\documentclass[9pt]{article}  
\usepackage{times}
\usepackage{graphicx}

\usepackage{color}

\usepackage[round,numbers,sort&compress]{natbib} 

\usepackage{amsmath}
\usepackage{amssymb}
\usepackage{stackrel}
\usepackage{chemarrow}
\usepackage{graphicx}
\usepackage{textgreek}

\usepackage{booktabs}
\usepackage{dcolumn}
\usepackage{rotating}
\usepackage{multirow}
\usepackage[colorlinks, linkcolor = black, citecolor = black, filecolor = black, urlcolor = blue]{hyperref}

\usepackage[british]{babel}
\usepackage{stfloats}

\renewcommand\appendix{\par
  \setcounter{section}{0}
  \setcounter{subsection}{0}
  \setcounter{figure}{0}
  \setcounter{table}{0}
  \renewcommand\thesection{Appendix \Alph{section}}
  \renewcommand\thefigure{\Alph{section}\arabic{figure}}
  \renewcommand\thetable{\Alph{section}\arabic{table}}
}

\begin{document}

\twocolumn[{\LARGE \textbf{Comment on Tamagawa \& Ikeda's reinterpretation of the Goldman-Hodgkin-Katz equation}\\*[0.1cm]
{\large Are transmembrane potentials caused by polarization?\\*[0.2cm]}}
{\large Thomas Heimburg\\*[0.1cm]
{\small Niels Bohr Institute, University of Copenhagen, Blegdamsvej 17, 2100 Copenhagen \O, Denmark - theimbu@nbi.ku.dk}\\*[-0.1cm]

{\normalsize \textbf{ABSTRACT}\hspace{0.5cm} The emergence of electrical fields across biological membranes is central to our present understanding of biomembrane function. The most prominent example is the textbook model for the action potential \cite[]{Hodgkin1952b} that relies on transmembrane voltage and membrane permeability. In a recent article by Tamagawa \& Ikeda (\href{https://link.springer.com/article/10.1007/s00249-018-1332-0}{$\rightarrow$link}), an important underlying concept, the Goldman-Hodgkin-Katz equation, has been challenged \cite{Tamagawa2018}. This will be discussed below.
\\*[0.3cm] }}
]

In the view of electrophysiology, the transmembrane vol\-tage is the consequence of the existence of semi-permeable walls, i.e., a membrane that in some regions is only permeable to a particular ion, e.g., potassium, and impermeable to other ions. Other regions of the membrane may display selectivity for different ions, e.g., sodium. The Nernst-Planck equation considers the transmembrane ion current as the sum of the charge flow in the electrical field, and the diffusive currents due to ion concentration differences \cite[]{Goldman1943}. It is proportional to the membrane permeability, $P$. The steady state solution in the simplest case of permeability for just one single ion is the Nernst potential,
\small
\begin{equation}\label{eq.01}
V=\frac{RT}{zF}\ln \frac{[C]_{out}}{[C]_{in}}\;,
\end{equation}
\normalsize
where $[C_{in}]$ and $[C_{out}]$ are inside and outside ion concentrations,  and $z$ the valency of the respective ion. This is the equilibrium case where diffusive and electric forces are balanced, and the current across the membrane is zero. The Nernst potential is a function of the ratio between inside and outside concentrations but does not depend on the permeability.

In a biological cell, many ions are present simultaneously. Typically, no voltage will exist for which the currents of all ions are zero simultaneously. This case is addressed by the Goldman-Hodgkin-Katz (GHK) equation, which is derived from the Nernst-Planck equation. It determines a voltage called the resting potential, $V_R$, at which the sum of all ion currents is zero. For instance, in the presence of potassium, sodium and chloride ions one finds
\small
\begin{equation}\label{eq.02}
V_R=\frac{RT}{F}\ln\left(\frac{P_K [K]_{out}+P_{Na} [Na]_{out}+P_{Cl} [Cl]_{in}}{P_K [K]_{in}+P_{Na} [Na]_{in}+P_{Cl} [Cl]_{out}}\right) \;,
\end{equation}
\normalsize
where the $P_i$ are the permeabilities for the different ions. This equation describes a non-equilibrium situation where permeabilities matter. Individual currents exist, but the sum of all currents is zero. For this reason, the resting potential has to be maintained by the activity of ion pumps. If the membrane is only permeable to one single ion (e.g., $P_{Na}=P_{Cl}=0$) one recovers the respective Nernst potential. The GHK equation provides an understanding for rectification, i.e., non-linear current-voltage relations that are a central element in the Hodgkin-Huxley (HH) model \cite[]{Hodgkin1952b}. Further, it predicts the direction of currents, e.g., after a temporary voltage excitation. Therefore, the theoretical concepts underlying the GHK-equation are very important for the present interpretation of electrophysiological data. 

Today, it is generally believed that specific membrane permeabilities for different ions are controlled by voltage-gated ion channel proteins, and that the conductances of the specific channels are proportional to the macroscopic membrane permeability for the respective ions. The notion of selective ion channel proteins is plausible because different phases of the transmembrane current can be attenuated by the use of toxins such as tetrodotoxin \cite[]{Moczydlowski2013}, which is attributed to the blocking of individual channels by specific binding. However, it is hard to detect ion flux and selectivity on a molecular scale.\\

\noindent\textbf{Another interpretation of the GHK equation: }
Recently, Tamagawa and Ikeda suggested that the GHK-equation can be interpreted in a completely different way, namely by an ion adsorption mechanism rather than a permeation process \cite[]{Tamagawa2018}. They attribute this idea to the American cell physiologist Gilbert Ling \cite[]{Ling1962, Ling2001}, who is an opponent of the concept of ion channels and pumps. While controversial, Ling's concepts found numerous supporters because they are solidly rooted in physical chemistry and thermodynamics.

Tamagawa and Ikeda  reported experiments with two glass beakers, one containing KCl and the other one KBr. The two containers were electrically coupled by a wire and two AgCl electrodes, but no diffusion of ions between them could take place. This is equivalent to an impermeable wall. The authors demonstrate that the measured concentration dependence of the voltage follows the GHK-equation even though no ions can flow across the membrane. Tamagawa and Ikeda provide a simple theory. They assume that the impermeable wall possesses binding sites for ions with binding constant $K_i$ ($i$ is the index of an ion), and ions bind according to the mass action law. Therefore, the total amount of adsorbed ions depends on ion concentration. As a result, the surface acquires a net charge density. If the ion concentration on the other side of the impermeable wall is different, a transmembrane potential emerges that the authors calculate by using the well-known Gouy-Chapman theory, which relates surface charge density to surface potential. Surprisingly, the resulting equation is mathematically identical to the GHK-equation. In the case of potassium, sodium and chloride ions, they would obtain \cite[]{Tamagawa2018}):
\small
\begin{equation}\label{eq.03}
V_R=\frac{RT}{F}\ln\left(\frac{K_K [K]_{out}+K_{Na} [Na]_{out}+K_{Cl} [Cl]_{in}}{K_K [K]_{in}+K_{Na} [Na]_{in}+K_{Cl} [Cl]_{out}}\right) ,
\end{equation}
\normalsize
which is identical to eq. (\ref{eq.02}), but replaces permeabilities, $P_i$, by adsorption constants, $K_i$. The difference between the two interpretations of this equation is that eq. (\ref{eq.02}) describes a steady state situation that is not at equilibrium, while eq. (\ref{eq.03}) refers to an equilibrium state where no ions flow and no ion pumps are required. This is one of the elements of Ling's association-induction hypothesis \cite{Ling1962}.

Tamagawa and Ikeda conclude from this that the GHK-equation is no reliable tool to determine permeabilities. In their view, the $P_i$ in the GHK-equation are merely parameters to fit the concentration dependence of the voltage, but they do not have to represent permeabilities.\\

\noindent\textbf{Polarization:} The differential adsorption of ions to the two sides of a membrane (described by Tamagawa and Ikeda) effectively creates a polarization in the absence of membrane permeability. It has in fact been shown that sodium and chloride ions can bind to lipid membranes \cite[]{Boeckmann2003}. There are other possible origins of polarization, e.g., an uneven distribution of charged lipids in both monolayers, asymmetric membrane proteins, or membrane curvature via the so-called flexoelectric effect \cite[]{Petrov1999, Petrov2001, Mosgaard2015a}. 
Although these effects clearly exist, polarization effects are totally absent in conventional biomembrane theories. When it comes to nerves, there exists an alternative approach to explain the nerve pulse, the so-called soliton model, which considers the action potential an electromechanical pulse \cite[]{Heimburg2005c}. Here, voltage changes are a consequence of variations in capacitance and polarization, and no selective permeabilities for ions are required.\\

\noindent\textbf{Conclusion: }The paper by Tamagawa and Ikeda addresses the contribution of binding-induced polarization to the electrical properties of membranes. Adsorption of ions to biomembranes is practically unavoidable. Therefore, it seems obvious that existing electrophysiological models are at least incomplete. In the opinion of Tamagawa and Ikeda, they may even be wrong. Channels and pumps may not be needed to understand transmembrane potentials, as suggested by Ling's association-induction hypothesis. 
Instead, Tamagawa and Ike\-da provide a refreshing reinterpretation of the electrical behavior of biomembranes, which is based on ion adsorption. I consider their remarkable results an important progress in the interpretation of electrical membrane phenomena from a physical-chemistry perspective.

\noindent\hrulefill
\vfill\null

\begin{thebibliography}{10}
	
	\bibitem{Hodgkin1952b}
	Hodgkin, A.~L., and A.~F. Huxley.
	\newblock 1952.
	\newblock A quantitative description of membrane current and its application to
	conduction and excitation in nerve.
	\newblock J.\ Physiol.\ London 117:500--544.
	
	\bibitem{Tamagawa2018}
	Tamagawa, H., and K.~Ikeda.
	\newblock 2018.
	\newblock Another interpretation of the {G}oldman-{H}odgkin-{K}atz equation
	based on the ling's adsorption theory.
	\newblock Eur.\ Biophys.\ J. https://doi.org/10.1007/s00249-018-1332-0.
	
	\bibitem{Goldman1943}
	Goldman, D.~E.
	\newblock 1943.
	\newblock Potential, impedance and rectification in membranes.
	\newblock J.\ Gen.\ Physiol. 27:37--60.
	
	\bibitem{Moczydlowski2013}
	Moczydlowski, E.~G.
	\newblock 2013.
	\newblock The molecular mystique of tetrodotoxin.
	\newblock Toxicon 63:165--183.
	
	\bibitem{Ling1962}
	Ling, G.~N., 1962.
	\newblock A physical theory of the living state: the association-induction
	hypothesis.
	\newblock Blaisdell Publishing Company, New York, London.
	
	\bibitem{Ling2001}
	Ling, G.~N., 2001.
	\newblock Life at the Cell and Below-Cell Level. The Hidden History of a
	Fundamental Revolution in Biology.
	\newblock Pacific Press.
	
	\bibitem{Boeckmann2003}
	B{\"o}ckmann, R.~A., A.~Hac, T.~Heimburg, and H.~Grubm{\"u}ller.
	\newblock 2003.
	\newblock Effect of sodium chloride on a lipid bilayer.
	\newblock Biophys.\ J. 858:1647--1655.
	
	\bibitem{Petrov1999}
	Petrov, A.~G., 1999.
	\newblock The lyotropic state of matter. Molecular physics and living matter
	physics.
	\newblock Gordon and {B}reach {S}cience {P}ublishers, Amsterdam.
	
	\bibitem{Petrov2001}
	Petrov, A.~G.
	\newblock 2001.
	\newblock Flexoelectricity of model and living membranes.
	\newblock Biochim.\ Biophys.\ Acta 1561:1--25.
	
	\bibitem{Mosgaard2015a}
	Mosgaard, L.~D., K.~A. Zecchi, and T.~Heimburg.
	\newblock 2015.
	\newblock Mechano-capacitive properties of polarized membranes.
	\newblock Soft Matter 11:7899--7910.
	
	\bibitem{Heimburg2005c}
	Heimburg, T., and A.~D. Jackson.
	\newblock 2005.
	\newblock On soliton propagation in biomembranes and nerves.
	\newblock Proc.\ Natl.\ Acad.\ Sci.\ USA 102:9790--9795.
	
\end{thebibliography}


\small{

}

\vspace{4.8cm}


\noindent\hrulefill
\end{document}